# SELECTIVE REFLECTION SPECTROSCOPY AT THE INTERFACE BETWEEN A CALCIUM FLUORIDE WINDOW AND Cs VAPOUR

*Athanasios Laliotis, Isabelle Maurin, Michèle Fichet, Daniel Bloch, Martial Ducloy*

*Laboratoire de Physique des Lasers UMR 7538 du CNRS, Université Paris-13, F-93430 Villetaneuse, France*

*Nikolai Balasanyan, Apetnak Sarkisyan, David Sarkisyan*

*Institute for Physical Research, NAS of Armenia, Ashtarak-2, Armenia*

<u>e-mail</u> : *bloch@lpl.univ-paris13.fr*

*A special vapour cell has been built, that allows the measurement of the atom-surface van der Waals interaction exerted by a $CaF_2$ window at the interface with Cs vapour. Mechanical and thermal fragility of fluoride windows make common designs of vapour cells unpractical, so that we have developed an all-sapphire sealed cell with an internal $CaF_2$ window. Although impurities were accidentally introduced when filling-up the prototype cell, leading to a line-broadening and shift, the selective reflection spectrum on the Cs $D_1$ line (894 nm) makes apparent the weak van der Waals surface interaction. The uncertainties introduced by the effects of these impurities in the van der Waals measurement are nearly eliminated when comparing the selective reflection signal at the $CaF_2$ interface of interest, and at a sapphire window of the same cell. The ratio of the interaction respectively exerted by a sapphire interface and a $CaF_2$ interface is found to be 0.55 ± 0.25, in good agreement with the theoretical evaluation of ~0.67.*

1. Introduction

The long-range van der Waals (vW) atom-surface attraction [1] tackles various fundamental problems, and imposes various constraints in the development of nanotechnologies, like the deposition of atoms in nano-lithography. With regards to its universality and to the possibilities to achieve exotic situations [2-4], the experimental investigations have remained relatively limited, and in particular, the list of materials submitted to experimental investigation has remained short. Most of the experiments have employed an optical detection, hence requiring transparency of the surface (occasionally, a thin metal film provides a sufficient transparency [1,5]). To investigate these vW effects, we have developed for a long time Selective Reflection (SR) spectroscopy at a vapour interface [1,2,6-9], a technique that typically probes the vapour at the interface on a depth $\lambda/2\pi$, with $\lambda$ the optical irradiation wavelength, and more recently extended it to spectroscopy in a vapour nanocell [10]. Materials were limited to glass –fused silica- , sapphire and YAG (for nanocells, to YAG and sapphire only). Alternate measurements were performed with cold atoms, at a glass interface or at an interface of a glass piece coated with a thin layer (see *e.g.* [10] for references). Extending the number of materials for the measurement of the vW interaction is of interest, especially when exotic effects such as vW repulsion [2], occur for specific interfaces [2-4, 11].

A particular interest of fluoride crystals, is that their surface resonances are located in the thermal IR range. This makes them good candidates to investigate how the thermal emissivity of the surface influences the vW interaction exerted onto a resonantly absorbing atom [3,4,11]. To implement the corresponding SR experiments, that require both a relatively dense (hot) vapour of alkali (Cs in our case), and hot windows of a controllable temperature, a specially designed vapour cell must be built-up. Indeed, fluoride materials are mechanically fragile and exhibit a large thermal expansion coefficient, making unpractical most of the standard designs for glass or sapphire cells. In the present work, a prototype Cs vapour cell with a $CaF_2$ window was constructed (section 2). A SR experiment on the $D_1$ resonance line (894 nm) was implemented (section 3), in order to test the cell *via* the evaluation (section 4) of the vW atom-surface interaction. Impurities having appeared when filling-up the cell make the measurement more complex, but we show that the vW interaction exerted by the $CaF_2$ window can be reliably compared to the one exerted by a sapphire interface when recording and fitting simultaneously the two spectra.

## 2. Design of a cell with a $CaF_2$ window

The vapour cell must obey the following requirements: it must resist to a high temperature (to warrant a sufficient density of Cs and for future experiments involving thermal emissivity), it must be evacuated to avoid unwanted collision broadening and shift, and it has to be compatible with a $CaF_2$ window. We discuss below the available technologies before describing our cell.

The current technology with glass-blowing allows the convenient build-up of temperature-resistant (~ 200 °C) sealed glass cells, but it cannot be extended to the sealing of special crystalline windows. Sealed cells with a metallic housing and various windows such as sapphire or even fluoride windows, can be prepared with commercial systems of windows soldered to a metallic mount; however, it is advised for the metal/window region not to exceed a maximal temperature of ~200°C. Alternately, a metallic housing cell, with windows mechanically hold by metal (copper) rings, is conceivable for a system under permanent evacuation but this makes difficult to monitor the vapour pressure. Such a design, operative with materials of a sufficient mechanical resistance, is at a high risk with fragile fluoride windows.

All-sapphire cells [12] can be realized with a special mineral glue connecting a sapphire tube and sapphire windows, the cell being sealed after evacuation and introduction of the alkali-metal. Such cells resist to very high temperature (up to ~1000 °C with an appropriate cell design). If sapphire windows can easily be replaced by nonbirefringent YAG windows, such a replacement is not permitted for fluoride materials as due to their mechanical fragility and to very different thermal expansion coefficients. In a similar manner, the mineral gluing technology cannot be extended to an all-fluoride cell.

This led us to design our special cell on the basis of an all-sapphire sealed cell, with a fluoride window inserted inside the tubing, mechanically maintained close to one of the window. An issue for the observation of a SR signal at the $CaF_2$ interface is to make negligible the absorption in this interstitial region.

The cell (fig. 1) has a T-shaped geometry, in order to allow for an independent control of the vapour density (through the temperature of the alkali-metal reservoir), and of the temperature of the window (controlled through the temperature of the cell body). It has two opposite windows, one of which is in YAG, and the other one in sapphire ($\perp$ c-axis, to avoid birefringence). The $CaF_2$ window consists of a long (~ 7 cm in our prototype cell) cylindrical piece. This allows the $CaF_2$ tube to accommodate a temperature gradient: the YAG window is maintained at room temperature outside the oven, so that the corresponding extremity of the $CaF_2$ tube remains cold, while the other extremity of the $CaF_2$ tube reaches an adjustable temperature (up to ~350 °C) (fig. 1). This makes the free-space between the $CaF_2$ window and the YAG window non critical: indeed, at room temperature, the very low alkali-metal vapour density yields negligible effects on the transmission (conversely, in the absence of a temperature gradient, the SR signal at the $CaF_2$ interface is contaminated by the transmission signal of the interstitial volume for separation distances as short as 100 nm). To reduce still this vapour density, and to avoid Cs condensation in the interstitial region, a cooled (~ 5°C) metal ring is usually set around the sapphire tubing.

As shown in fig. 1, the oven has a complex structure and is made of independent sections, each of them contains heating wires (structured in doubled coils to minimize the induced magnetic field). These wires are surrounded by thermally isolating bricks of alumina powder. Note that, as usual in the construction of a T-shaped all-sapphire cell, a convenient filling-up of the cell reservoir imposes the bottom part of the reservoir to be in glass (and hence at $T < 200°C$) but the Cs density is governed by the temperature of the upper free-surface, *i.e.* the temperature of the oven $T_{res}$. Also, the T-shape of the cell divides the heating wire in the top part of the oven ($T_{windows}$) in two independent subsections. This is the reason for a temperature gradient inside the top part of the oven (the temperature of the top part is monitored through two thermal probes located close to the two respective windows), that can be suppressed if sending independent currents in the heating wires, or if specifically adjusting the lengths of the wire. In view of experiments where the surface temperature could be varied while keeping constant the Cs density (in principle governed by the temperature in the Cs reservoir), the independence of the top and bottom sections of the oven was also tested. An adjustable temperature difference 30-120°C is currently obtained for a bottom part at 150-200°C (we operate at a minimal ~30°C nominal difference because of the uncertainty in the temperature measurement). With an improved isolation between the oven sections, it should be not difficult to increase this relative independence to a larger scale, and the upper limit in temperature is essentially a matter of the outgassing conditions.

Although we could not find any data on the chemical resistance of fluoride windows to hot alkali vapour, and in spite of the possibility of a migration of fluor atoms from heated fluoride materials, no special problems have been detected when repeatedly heating the cell up to ~350 °C. Unfortunately, in spite of a standard procedure for the cell outgassing (350°C**,** 10 hours), an incident has apparently occurred when filling-up the cell with Cs, that brought impurities in the only cell prototype available for a long duration. These impurities are responsible for a major (collision) broadening of the Saturated Absorption (SA) signal observed on the $D_1$ line of Cs (894 nm), turning it into a Doppler-broadened signal instead of a Doppler-free one. The expected sub-Doppler signal, that is nevertheless tiny and appearing on a broad background (figure 2), is recovered only with a fast modulation [13], that minors the influence of velocity-changing collision processes. As shown below, these impurities have a less dramatic effects in SR spectroscopy.

### 3. Experimental set-up and main spectroscopic features

The smaller sensitivity of SR spectroscopy to impurities, as compared to the SA technique, originates in its linear nature, and its fast response time (solely governed by the optical transition, and not by the build-up of an optical population). However, one still clearly observes a residual broadening in our SR experiments.

Several SR spectroscopy experiments were already performed on the Cs resonance lines, allowing to measure the vW interaction. The present one, schematized in figure 3, does not differ essentially from these previous experiments. Rather, a double SR set-up has been implemented in order to monitor simultaneously the SR spectrum at the $CaF_2$ interface, and at the opposite sapphire window. The laser is here a tuneable narrow-linewidth DBR laser at 894 nm, and its frequency scan linearity is monitored by analyzing the transmission through a low-finesse very stable Fabry-Perot (FP). The laser is also frequency modulated (FM), in order to obtain purely Doppler-free SR spectra [14] through demodulation of the reflected beam. A frequency marker for the atomic resonances in volume is usually required for the analysis of the vW interaction in the SR spectrum: this is why a simultaneous SA experiment is implemented in an auxiliary cell, with an amplitude modulation (AM) applied to the pump beam. Note that to compare the atomic resonances in volume, and close to the surface, the pressure conditions should be the same as in the SR cell. This is why the tiny and narrow SA signal available in the special cell when the pump undergoes a fast modulation, was compared to the SA signal of the auxiliary experiment. The SA signal in the special cell appears red shifted by ~ - 5.2 ± 0.5 MHz for all h.f.s. components (experiments performed at room temperature so that the pressure shift induced by Cs-Cs collisions is entirely negligible).

To evaluate the possible competitive presence of a spectroscopic signal originating in the thin interstitial region between the $CaF_2$ window and the YAG window, we have attempted to measure the (FM) signal reflected by the $CaF_2$ - YAG interface. The SR signal, typically integrating the atomic response over $\lambda/2\pi$, is hence negligible relatively to the Doppler-broadened linear absorption signal resulting from the back and forth propagation through the interstitial region. Comparing this signal with the absorption through the whole cell kept at room temperature (vapour column ~8 cm), we extrapolate that the interstitial spacing is on the order of 50 µm (the back and forth absorption in this interstitial region is below $10^{-3}$, and is observable only through a large amplitude FM). This ensures that when recording the SR spectrum with a hot vapour, the signal of the transmission through the interstitial region remains largely negligible: the contribution of the interstitial region is still reduced by the cooling close to the YAG window, and by the FM technique that specifically enhances the narrow signals.

We have recorded the SR spectra on the Cs $D_1$ line $6S_{1/2}(F=3,4) \rightarrow 6P_{1/2}(F'=3,4)$ across the four (entirely resolved) h.f.s. components. The signals from the sapphire and from the $CaF_2$ windows were recorded simultaneously, and for various temperatures of the Cs reservoir. The irradiation intensity is kept low ( ≤ 0.1 mW/cm² ) to ensure the linearity of the SR technique. The temperature of the top part of the cell has no significant influence over the SR signal amplitude. At low temperatures (~ 100°C), when the Cs self-broadening is negligible, the minimal SR width is found to be on the order of 27 MHz, largely exceeding the natural width of the transition. The SR spectra exhibit a notable asymmetry, along with a red frequency shift, typical of the vW interaction (see fig.4). The apparent shift, relatively to the SA frequency marker, varies slightly with the Cs pressure conditions and with the considered h.f.s. components. It is actually much larger than the red shift currently found for the vW interaction, on the order of 2 -3 MHz, and hence originates in the residual impurities in the cell, and possibly in a Cs-pressure shift for the higher Cs temperatures. As shown in the next section, it remains possible to extract an estimate of the vW interaction, although the vW effects are largely hindered by volume collisions effects.

## 4. Analysis of the vW interaction

The vW interaction simultaneously shifts and distorts (asymmetry) the optical transition. The vW interaction on the $D_1$ resonance line is known to perturb only weakly the (FM) SR spectrum, and is responsible for an approximate spectral shift not exceeding a few MHz, smaller than the minimal width of the transition (~ 5 MHz in the absence of any collisions). It had however been demonstrated that, even when collisions (buffer gas, or self-broadening) impose the dominant broadening, the vW interaction, although of a weaker influence, had to be approximately taken into account to allow for a correct evaluation of the influence of collisions [8]. Here, the novelty is that we aim at evaluating as precisely as possible the vW interaction exerted by the $CaF_2$ window, or at least to compare it with the one exerted onto the sapphire window, in spite of the large impurities broadening.

We recall here the general method for extracting the vW interaction from a (FM) SR spectrum. When the Doppler width largely exceeds the optical width, and in the linear regime, (FM) SR experimental lineshapes are amenable to dimensionless lineshapes [1,6-9], dependent on a single dimensionless parameter A characterizing the vW interaction ($A = 2C_3k^3/\gamma$, with $k=2\pi/\lambda$, $\gamma$ the optical width of the relevant transition [6], and $C_3$ the coefficient of the spatial dependence of the transition frequency, evolving according to $\omega(z) = \omega_\infty - C_3 z^{-3}$, with z the atom-surface distance). In numerous realistic cases, the finite ratio between the Doppler width (ku) and the optical width ($\gamma$) imposes a slight correction to the dimensionless theoretical lineshapes [1], but no additional parameter is introduced as the Doppler width is intrinsically fixed by the experimental conditions. One hence attempts to fit an experimental lineshape with theoretical dimensionless lineshapes for different A values. The adjustable parameters are the amplitude, width, location of the resonance (when not imposed by the free-space reference), and possibly a vertical offset. $C_3$ is hence determined from the limited range of A values allowing an acceptable fitting. The consistency of the fittings is usually checked by establishing that a single $C_3$ value is found in spite of phenomenological changes imposed to the lineshapes when varying the pressure (see *e.g.* [1,2,7-9]).

In the "weak vW regime" ($A \ll 1$) of our experiments, the optimal fitting width $\gamma(A)$ is nearly independent of the A value, but there is a strong interplay between the location of the resonance (*i.e.* frequency shift as possibly resulting from collisions) and the A value (fig. 4). This makes it very difficult to extract unambiguously the VW interaction from a single SR spectrum, notably when the spurious shift originating in the residual impurities is unknown. This is why we have attempted to benefit from the simultaneous recording of the SR signal at the two different interfaces, assuming that the Cs vapour (density, and collisions) is homogeneous in the top part of the cell. Hence, only a single parameter for the width, and another one for the shift, are to be adjusted when simultaneously fitting the experiments with sapphire and with $CaF_2$. The tighter constraint imposed in the fitting method bring a serious help to the consistency of the fittings. In addition, the coupled fitting of two independent signals recorded simultaneously with the same laser eliminates most of the effects related to a possible tiny laser frequency drift.

In our situation, the vW coefficients for $CaF_2$ and sapphire are expected to differ notably, owing to an approximate $(\varepsilon-1)/(\varepsilon+1)$ dependence [1], yielding a theoretical ratio ~ 0.67 (as predicted from the respective refractive indices of $CaF_2$ (n=1.430) and sapphire (n=1.758), and $\varepsilon = n^2$). Hence, in analyzing the data, the objective is to find the optimal range of coefficients $(A_1, A_2)$ characterizing the vW coupling. Figure 5 shows that high quality fittings of the experimental curves can be obtained in spite of the restricted number of adjustable parameters. Significantly, for each experimental curve to be fitted, the domain of acceptable $(A_1, A_2)$ values is strongly limited, and the cloud of fitting $(A_1, A_2)$ parameters is notably characterized by the ratio $A_1/A_2$ between the interaction exerted by the $CaF_2$ window and by the sapphire window. In addition, the assumption that the Cs vapour density is the same at the two interfaces should impose the relative amplitudes of the two SR signals. This is indeed verified, within the limits of the experimental uncertainty, mostly affecting the non resonant reflection, which usually departs from the predicted value (*i.e.* Fresnel formula) owing to scattering losses at the interface. As illustrated in figure 6, the estimated $C_3^{sapphire}/C_3^{CaF2}$ ratio remains constant (~ $0.55 \pm 0.25$) when varying the temperature conditions, in very good agreement with the theoretical value (~0.67). The accuracy for the individual values for $C_3$ is lower than for the relative interaction exerted by sapphire and $CaF_2$ respectively; the obtained values $C_3^{sapphire} = 1.4$ kHz.µm$^3$, $C_3^{CaF2} = 1.0$ kHz.µm$^3$ are in an acceptable agreement with the predicted ones (respectively 1.0 kHz.µm$^3$, and 0.67 kHz.µm$^3$, following a 2.0 kHz.µm$^3$ value predicted for the $D_1$ line at an ideal interface, according to $C_3^{ideal}(6P_{1/2}) = 4.43$ kHz.µm$^3$ [2], and $C_3^{ideal}(6S_{1/2}) = 2.42$ kHz.µm$^3$ [7]). At last, an optical width and collision shift are associated to each fitting of an experimental recording, allowing to plot a Cs pressure self-broadening and shift (figure 7). These variations of the pressure effects, partly dependent upon the investigated hyperfine component, are compatible with previous estimates of self-broadening and shift [15,16], although the accuracy, affected as usual by the pressure scale uncertainty (~ 40 % originating in a ~5°C temperature uncertainty), suffers here from the offset introduced by the impurities broadening. Also, the residual shift and broadening attributed to the impurities are independent of the cell temperature, showing that the impurities that have appeared when filling-up the cell are of a nearly constant density, and not stored in a dense phase with a temperature-controlled gas equilibrium; moreover, the residual shift corroborates the SA experiment (section 3). As a further evidence of the consistency of our fittings and extrapolation from a series of spectra, fig. 8 shows the ability to fit a series of spectra with a single $C_3$ value (for respectively sapphire, and $CaF_2$), and the optical width and shift values deduced from fig. 7.

## 5. Conclusion

In summary, we have designed an alkali vapour cell with a fluoride window susceptible to be heated up to a high temperature. The principle of construction, allowing the contact of high temperature vapour with the window of interest, is actually applicable to any interface made of a crystal (or material) resistant to hot alkali vapour. In spite of spurious impurities that has degraded the vacuum conditions in the prototype cell, we show that the atom-surface vW interaction can be measured at the interface with the special window, hence validating the design of the construction. Such a cell is an important step to study the specific influence of the thermal emittance of fluoride windows on suitable atomic transitions, such as the Cs $6S_{1/2} \rightarrow 8P_{3/2}$ line [9]. In addition, it appears that the coupled analysis of the vW interaction exerted on two different windows, as extracted from a simultaneous recording, is a particularly robust method to estimate reliably the strength of the $C_3$ coefficient of the vW interaction, notably when stray effects can hinder the specific vW contribution.

**Acknowledgments**

This work was partially supported by FASTNet (European contract HPRN-CT-2002-00304) and by the INTAS South-Caucasus Project (grant: 06-1000017-9001).

**Figure captions**

Figure 1 : Scheme of the Cs vapour cell. In the top oven, the heating source is located in the left and right parts surrounding the connection to the Cs reservoir.

Figure 2 : Saturated Absorption (SA) spectrum recorded on the prototype cell with impurities (a) AM (500Hz) applied on the pump the signal, (b) In FM, the sub-Doppler signal emerges more clearly from the broad background for a fast FM.

Figure 3 : Scheme of the experimental set-up.

Figure 4 : An experimental SR spectrum ($T_{Cs}$ = 140 °C) recorded on the $CaF_2$ window. For A = 0.006 , A= 0.02, A = 0.04, the optimal fittings lead to the respective shifts  -7.1 MHz, -5.1 MHz, and -3.3 MHz, and widths 32.0 MHz, 32.6 MHz, and 32.9 MHz.  The vertical dashed line is the SA reference. The dotted line is a fit neglecting the vW interaction (A=0). The zoom on the left wing shows that it may be not easy to decide if A =0.02 is better than A = 0.006

Figure 5 : Optimal fitting (black) of two simultaneously recorded spectra (grey), under the constraint of a single  parameter, for the shift (-4.5 MHz), and respectively for the width (27.8 MHz). one has $T_{Cs}$ = 120 °C. The vertical dashed line is the SA reference. One has $A(CaF_2)$= 0.024 ($C_3$=0.96 kHz.µm$^3$), A(sapphire) = 0.042 ($C_3$=1.68 kHz.µm$^3$). The ratio of fitting amplitudes is 0.57 in the fitting,  0.516 in theory

Figure 6 : Ratio of the vW strength coupling $C_3$ between sapphire and $CaF_2$ window, as extrapolated from the fits for different Cs temperatures.

Figure 7 : Widths (a) and shifts (b) as a function of Cs pressure for the various hyperfine transitions, s extrapolated from the fitting of SR spectra.

Figure 8 : A series of experimental (grey) spectra fitted (black) with a unique $C_3$ value (1.4 kHz/µm$^3$ for sapphire, 1 kHz/µm$^3$ for $CaF_2$) and only adjustable amplitudes ; shifts and widths are also imposed as a result of the linear extrapolation of fig.7.

Figure 1

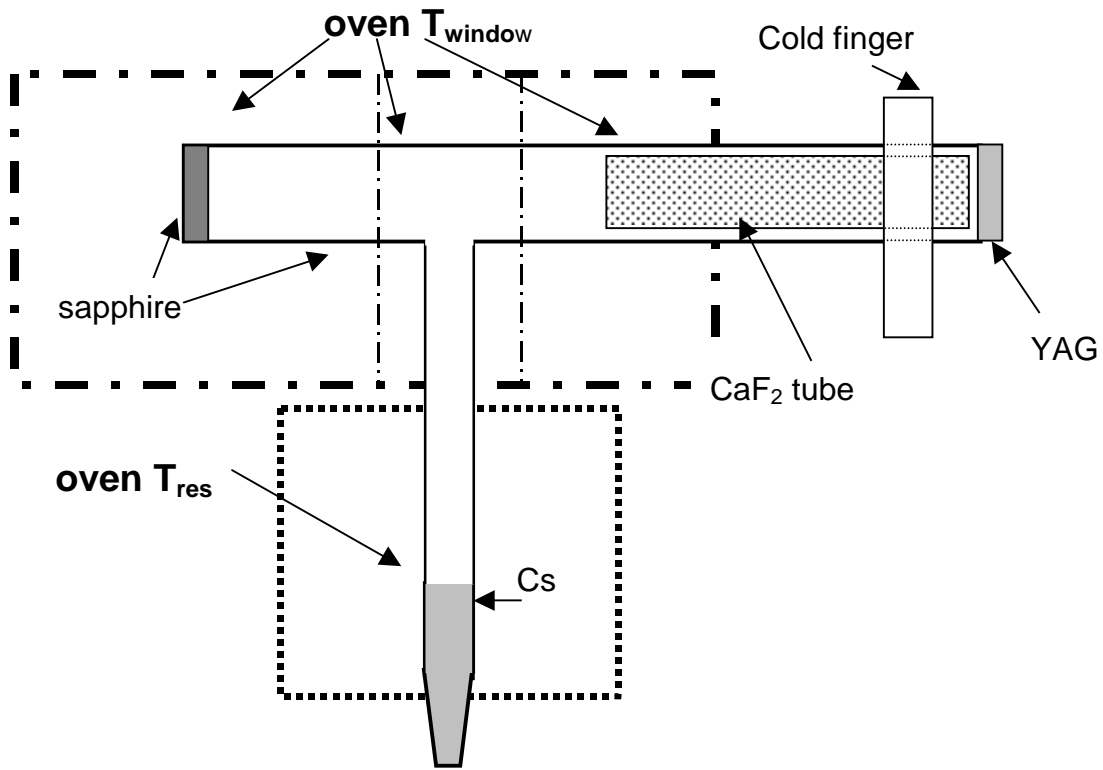

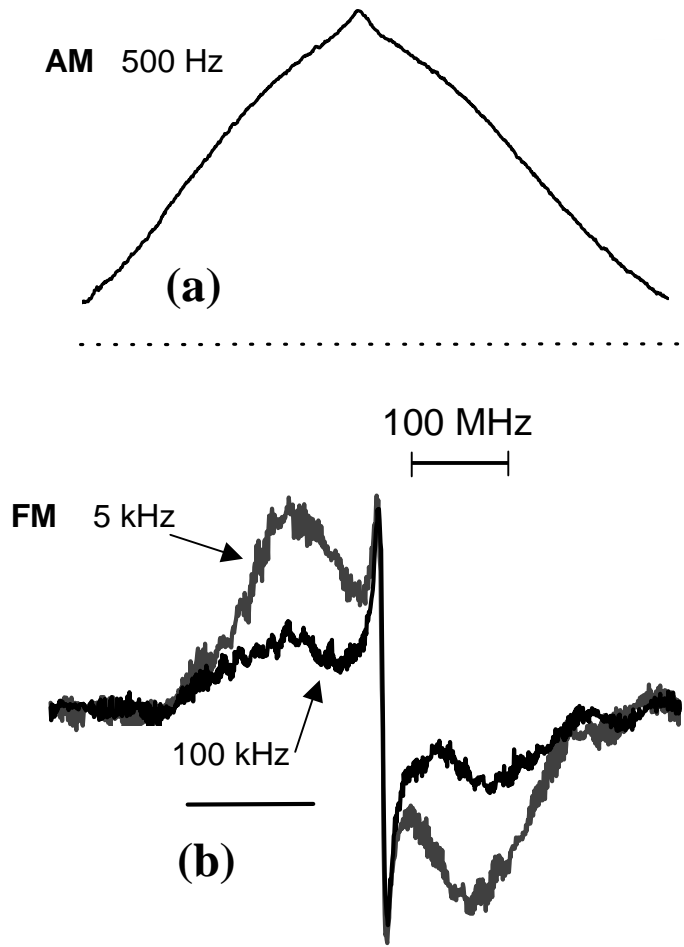

Figure 2

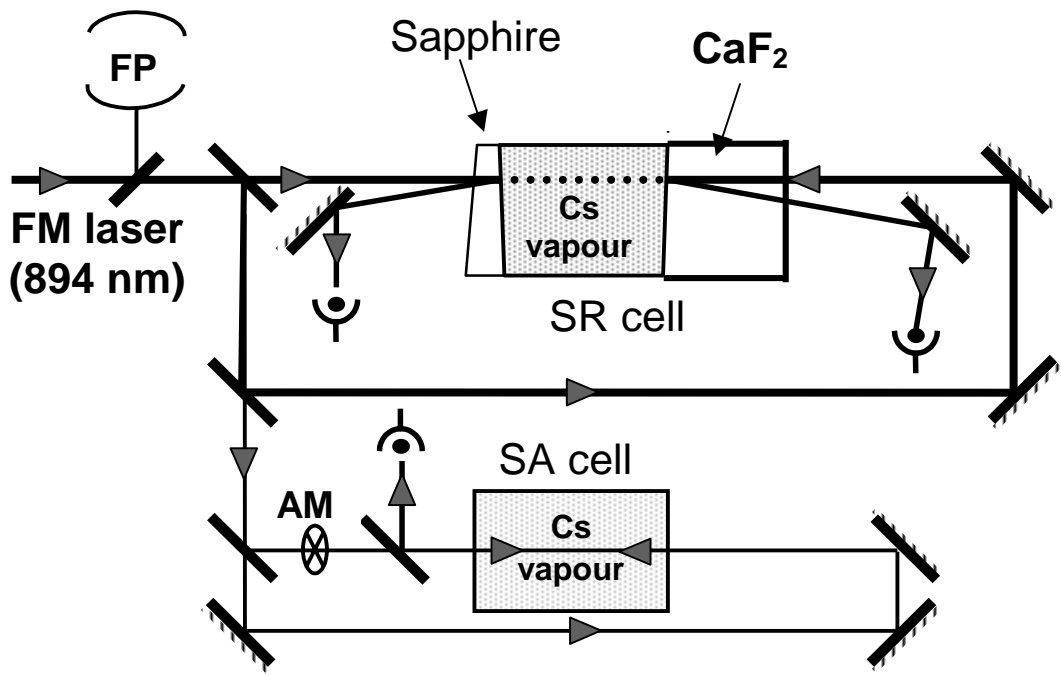

Figure 3

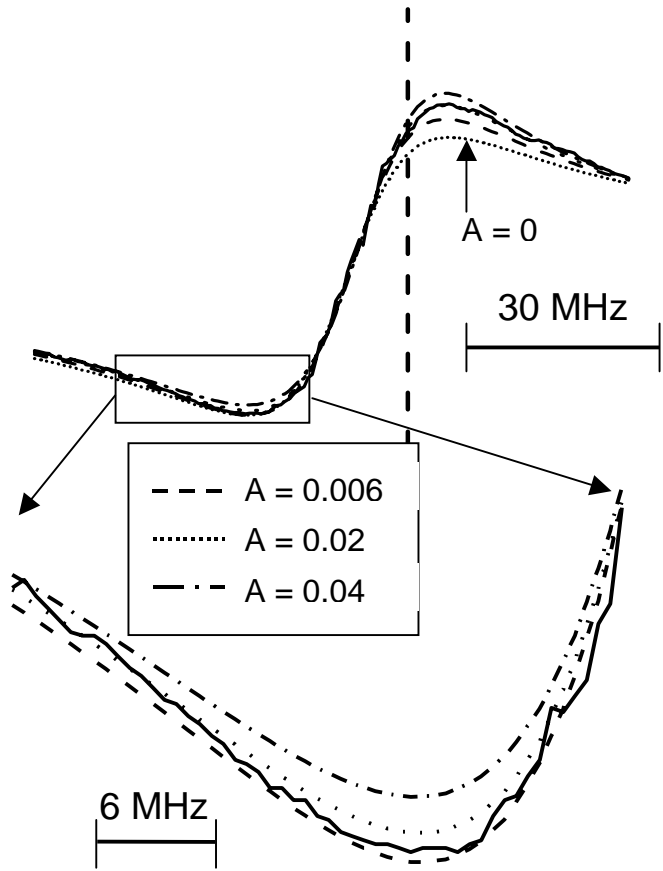

Figure 4

Figure 5

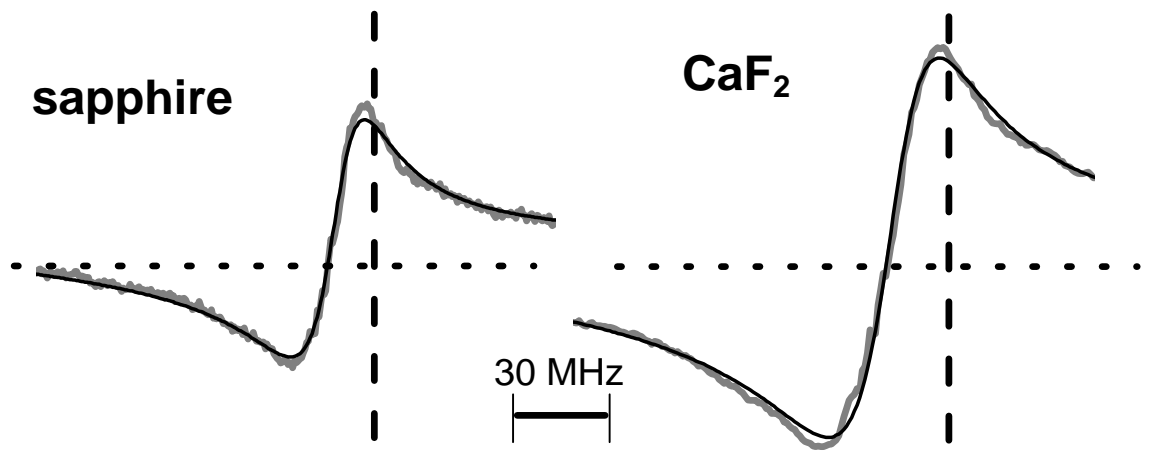

Figure 5

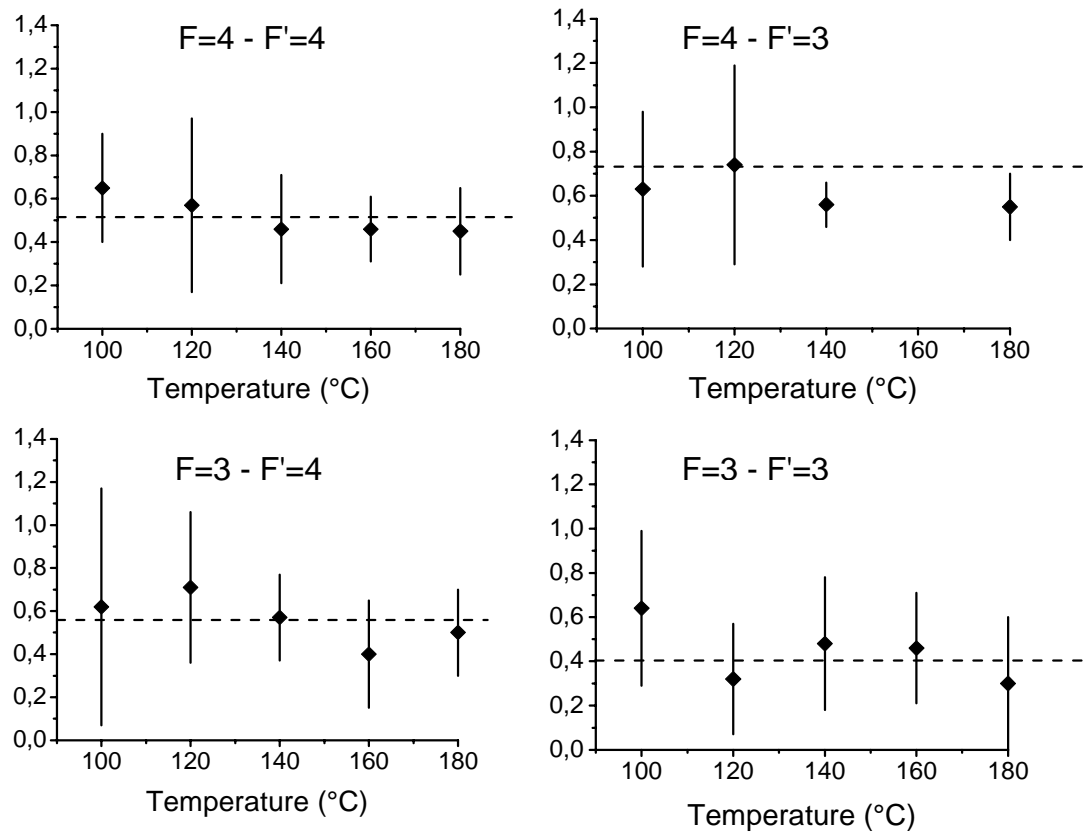

Figure 6

Laliotis *et al.*        Selective Reflection spectroscopy at the interface between a calcium fluoride window...

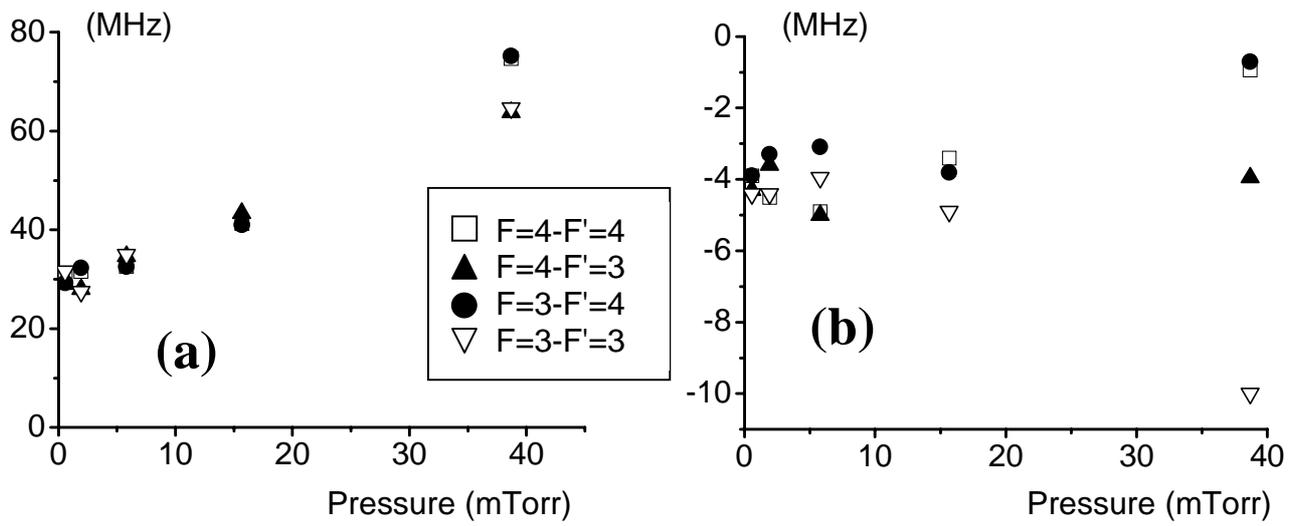

Figure 7

Figure 8

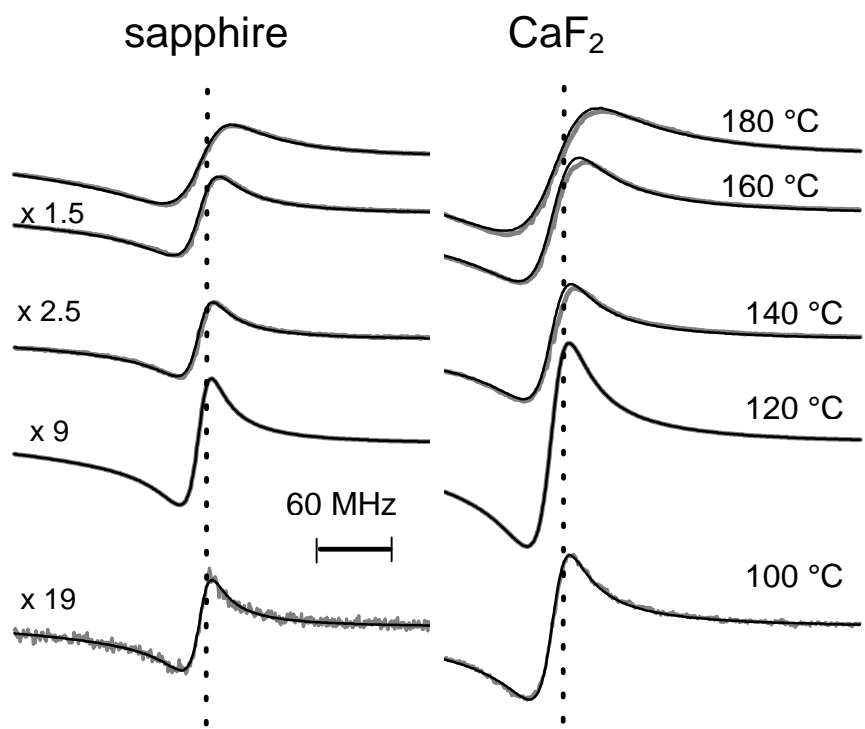